\documentclass[journal]{IEEEtran}

\usepackage{graphicx}
\usepackage{amsmath}
\usepackage[noadjust]{cite}
\graphicspath{ {images/} }
\usepackage[hyphens]{url}
\usepackage{hyperref}
\usepackage{eqexpl}
\eqexplSetDelim{ } 
  
\begin{document}

\title{The Creation and Validation of Load Time \\ Series for Synthetic Electric Power Systems}

\author{Hanyue Li,~\IEEEmembership{Student Member,~IEEE,}
        Ju Hee Yeo,~\IEEEmembership{Student Member,~IEEE,}
         \\
         Ashly L.Bornsheuer,~\IEEEmembership{Student Member,~IEEE,}
         Thomas J. Overbye,~\IEEEmembership{Fellow,~IEEE}
\thanks{The authors are with the Department
of Electrical and Computer Engineering, Texas A\&M University, College Station,
TX, 77840 USA (e-mail:hanyueli@tamu.edu, yeochee26@tamu.edu, abornsheuer@tamu.edu, overbye@tamu.edu).}%
}

%


\maketitle

\begin{abstract}

Synthetic power systems that imitate functional and statistical characteristics of the actual grid have been developed to promote researchers' access to public system models. Developing time series to represent different operating conditions of these synthetic systems will expand the potential of synthetic power systems applications. This paper proposes a methodology to create synthetic time series of bus-level load using publicly available data. Comprehensive validation metrics are provided to assure that the quality of synthetic time series data is sufficiently realistic. This paper also includes an example application in which the methodology is used to construct load scenarios for a 10,000-bus synthetic case.

\end{abstract}


\IEEEpeerreviewmaketitle

\section{Introduction}

\IEEEPARstart{P}{bulic} access to real power system data is limited due to confidentiality concerns. Synthetic power system models and data are created to be functionally and statistically similar to real power systems. Synthetic systems are synthesized using public data of the actual grid, and they don't represent the actual system located on the same geographic footprint, or contain any confidential information about the actual grid. 

Many efforts have been made on the creation of synthetic power system base cases, which contains systems topology and many of them have AC or DC power flow solutions. Early work of \cite{Zwang1}-\cite{Zwang2} came up with an approach to create transmission grid topologies based on the small world graph network.  A methodology for generating large scale synthetic transmission systems with AC power flow solutions on the footprint of United States was proposed in \cite{Adam1}-\cite{Adam2}, and several synthetic systems of different sizes and footprints were created. The work of \cite{Soltan1}-\cite{Soltan2} investigated the geographic and structural properties of North American and Mexican transmission grids, and created large electric systems with synthetic nodes and node connections. European synthetic transmission grids with DC power flow solutions were also developed based on public information from utilities and regulatory agencies in the synchronous grid of Continental Europe (UCTE) \cite{Bialek}.

Since synthetic power system base cases only reflect a one-time snapshot of the system, there is a natural need to expand the work and develop time series to represent changing system operating conditions over time. The combined data set of synthetic grid models and power system time series can be used as a benchmark for system scenario studies, a test bed for algorithms such as time sequence power flow, optimal power flow, and unit commitment.

Power system time series consist of data relevant to system status in a time sequence manner. It can span many aspects of the system such as the load, transmission line status, generator dispatch and electricity price. Real time series power system data are generally more publicly accessible compared to the data of actual grid topologies and models. For example, the Open Power System Data, for example, is a data depository that has load, wind, solar and price data in hourly resolution of 37 European countries \cite{OpenPowerSystemData}. In North America, for the transparent operation of electricity market, Independent System Operators (ISOs) often have hourly resolution load and price time series data publicly available as well \cite{ISO-NE}-\cite{IESO}. 

Besides actual system-level data, the creation of synthetic time series for household-level load is also a well-researched topic. The work of \cite{household1} developed a probabilistic mathematical model for residential load simulation. A top-down approach using domestic load patterns for household profile adaptation was implemented in \cite{household2}.  A machine learning method to generate synthetic residential building load time series from smart meter data was proposed in \cite{household3}. 
However, as most of the power system studies require load data resolution at the bus-level, either real system-level or synthetic household-level time series data can be directly applied to the synthetic grid.

This paper presents a methodology for synthesizing bus-level time series load data, building on the results of \cite{Scenarios}. The maximum value of load time series aims to match the corresponding load bus size determined in the above mentioned base case. The unique variation of each bus-level time series is a result of the heuristic aggregation of prototypical building and facility load time series.  To ensure the quality and realism of the synthetic load time series, comprehensive validation metrics are established from the actual system-level load time series. An example application using the created time series for synthetic electric grid scenario study is also given.

The creation and validation of load time series for  2,000 and  10,000- bus synthetic grids (i.e., the ACTIVSg2000 and ACTIVSg10K grids from \cite{tamurepository}) are used as examples. The ACTIVSg2000 synthetic system shares the same footprint as the Electric Reliability Council of Texas (ERCOT), and the ACTIVSg10K has the same service region as the Western Interconnection (WI).  However, the method of creating and validating synthetic time series is general enough to be applied to any system.

\section{Creation of Bus-level Load Time Series}
Electric load time series reflect electricity consumption patterns and provides insight on the absolute level and changing rate of load at different times. Having access to bus-level load time series is essential for the unit dispatch and commitment in power system operations since generation always needs to follow the time-varying system load. The load time series in synthetic power systems has hourly resolution with a duration of a year, and is created on the bus level so that every bus in the synthetic grid model has a unique profile. Each bus-level load time series is created using an iterative aggregation approach, where prototypical building load profiles are aggregated based on the size and composition of load buses. 

\subsection{Location and size of bus-level electric load}
The location and size of the electric loads are determined during the creation of the synthetic base case discussed in \cite{Adam1}-\cite{Adam2}. The load buses are located based on the clustering of geographic coordinates associated with postal codes that are obtained from the public U.S. census database. The size of each load bus is then scaled according to the population of the corresponding postal code and the per-capita MW consumption. A fixed power factor is assigned to each load as an assumption. 

The base case is used as a reference to create load time series. The size of load buses in the base case are considered to be the peak value of each bus-level time series, and the geographic coordinates assigned to each load bus are then used to determine the unique location-dependent load features such as load composition ratio and prototypical building load time series.

\subsection{Load bus composition ratio}
The assignment of a composition ratio of residential, commercial and industrial load on each bus is helpful to realistically represent the uniqueness of load. It establishes the geographic and demographic dependence of electric load similar to reality. 

 U.S. utility companies' service territories as well as their residential, commercial and industrial megawatt-hours sales values from the Annual Electric Power Industry Report are used to determine the bus load composition ratios \cite{AnnualElectricPowerIndustryReport}.  Each load bus is assigned to one utility company based off its geographic coordinates, and the company's sales ratio of the three load types is used as the average bus load composition ratio.   

\subsection{Prototypical building- and facility-level load time series}
To bridge between the bus load composition ratio, and a unique hourly profile, prototypical end user level load time series under residential, commercial and industrial load types are synthesized from public data. Building- and facility-level time series gives the desired bus load a good base to incorporate both individual user load patterns and the aggregation effect. Different categories of buildings and facilities and their prototypical load time series are realistic approximations to represent the most common and important load features.

\subsubsection{Prototypical residential and commercial building load}
The prototypical building load time series synthesized in this paper are the same as the ones developed in \cite{Scenarios}, where open source data of simulated hourly residential and commercial building energy consumption are used \cite{OpenEI}.

The residential data contains buildings' hourly electricity usage value from space heating/cooling, High Voltage Alternating Current (HVAC) fan, interior/exterior lighting, as well as appliances and miscellaneous loads. Each data file covers one typical meteorological year 3 (TMY3) location in the United States, which represents geographic locations with different meteorology\cite{OpenEIRes}. For commercial load, under each TMY3 location, 16 building electric load profiles are simulated using the Department of Energy (DOE) commercial reference building models\cite{OpenEICom}, and the contents in each time series data are like that of residential data set.

Under the United States footprint, 1020 residential and 16,320 commercial building time series are calculated as a summation of all electricity consumption categories under each building type. They are created to exhibit unique profiles of residential and commercial buildings, and electricity consuming variations over time and geographic location.

\begin{figure}[!h]
  \centering
  \includegraphics[scale = 0.4]{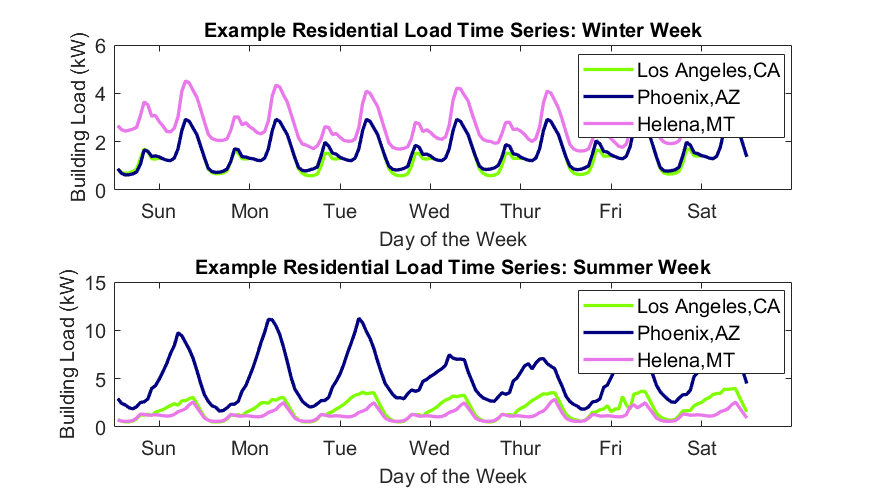}
  \caption{Prototypical residential building load time series examples by location}
  \label{fig:res_3tmy}
\end{figure}

Figure \ref{fig:res_3tmy}, for example, shows prototypical residential building load time series in a winter week and a summer week. The load shapes in two seasons are distinguishable, where winter profiles tend to have two peaks in one day due to winter heating, while summer profiles only have one peak per day. The magnitude of load can also be very different in each season, depending on geographic locations. In winter, regions with colder climate such as Helena, Montana,  would have higher average load. While in summer, load within hot and arid climate zones, such as Phoenix,Arizona, will have much more electricity consumption.  

\begin{figure}[!h]
  \centering
  \includegraphics[scale = 0.4]{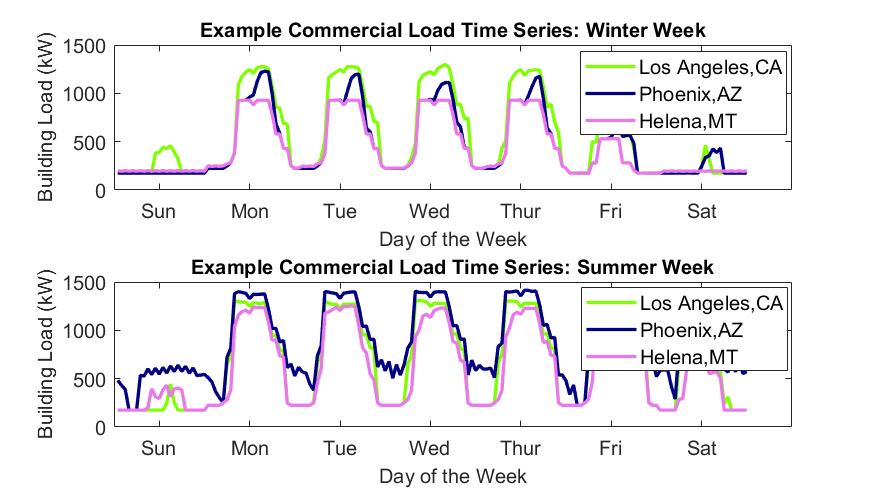}
  \caption{Prototypical commercial building load time series examples by location}
  \label{fig:com_3tmy}
\end{figure}

\begin{figure}[!h]
  \centering
  \includegraphics[scale = 0.4]{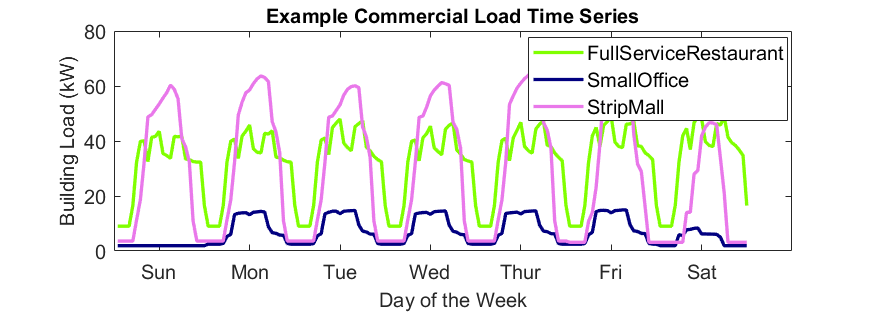}
  \caption{Prototypical commercial building load time series examples by type}
  \label{fig:com_3type}
\end{figure}

Similarly, figure \ref{fig:com_3tmy} shows prototypical commercial load time series for the large office building type. The load shape of a specific building type is generally consistent regardless of the location, and the load level is slightly higher in summer season compared to that in winter. In figure \ref{fig:com_3type}, weekly load profiles of three commercial building types (full-service restaurant, small office, and strip mall) in Los Angeles, California are shown. The load shape and size under each building type is unique. Small offices have steady load during weekdays and low load during weekends. For full-service restaurants and trip malls, load levels are constant through out the week, while full-service restaurants observe two peaks near lunch and dinner time, strip malls only peak once every day.

\subsubsection{Prototypical industrial facility load}
Prototypical industrial facility load time series are created based on publicly available per-unit industrial load curves from Oak Ridge National Laboratory (ORNL) \cite{ind} and the industrial assessment data in Industrial Assessment Centers (IAC) Database \cite{IAC}. 

The ORNL per-unit curves provide daily profiles of different industrial sectors, presented by different Standard Industrial Classification (SIC) codes with their unique load factor. The IAC Database contains information on the industry SIC code, total electricity usage and yearly operating hours of over 14,000 facilities in the United States, which are used to modify the ORNL curves into facility-specific load time series for a year. 

For each industrial facility, the yearly operating hour is first used to determine the total number of operating days. The ORNL daily curves of the same SIC code is then expanded to a yearly load curve,  with small white noise imposed and a random selection of starting day of the year. The synthesized yearly curve is then scaled so that the integral value of the curve matches the total electricity usage.

\begin{figure}[!h]
  \centering
  \includegraphics[scale = 0.4]{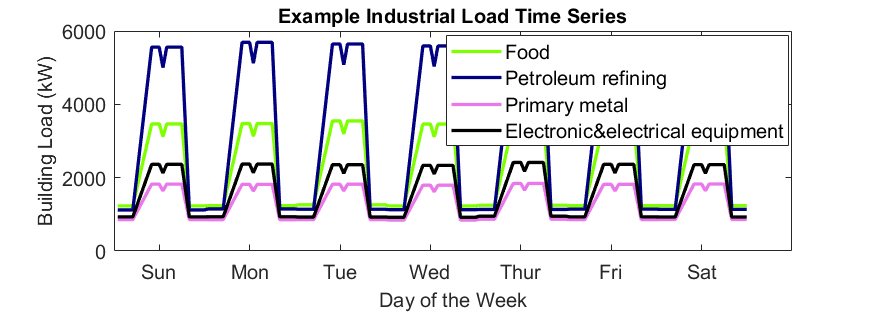}
  \caption{Prototypical industrial facility load time series examples by type}
  \label{fig:ind_4type}
\end{figure}

Figure \ref{fig:ind_4type} presents prototypical industrial load time series for four facilities from food, petroleum and refining, primary metal, as well as electronic and electrical equipment industries. As those load curves are adopted from the ORNL per-unit daily curves, they have similar daily variations and weekly shapes, with different load levels and load factors.

\subsection{Aggregation of load}
The bus-level load time series is created by iteratively aggregating prototypical building and facility load time series of each load type. This aggregation process has three main aspects: integrating realistic amounts of end users under each load type, selecting representative prototypical time series, and mimicking the effect of load aggregation described in \cite{AggregationEffect}.  

\begin{figure}[!h]
  \centering
  \includegraphics[scale = 0.44]{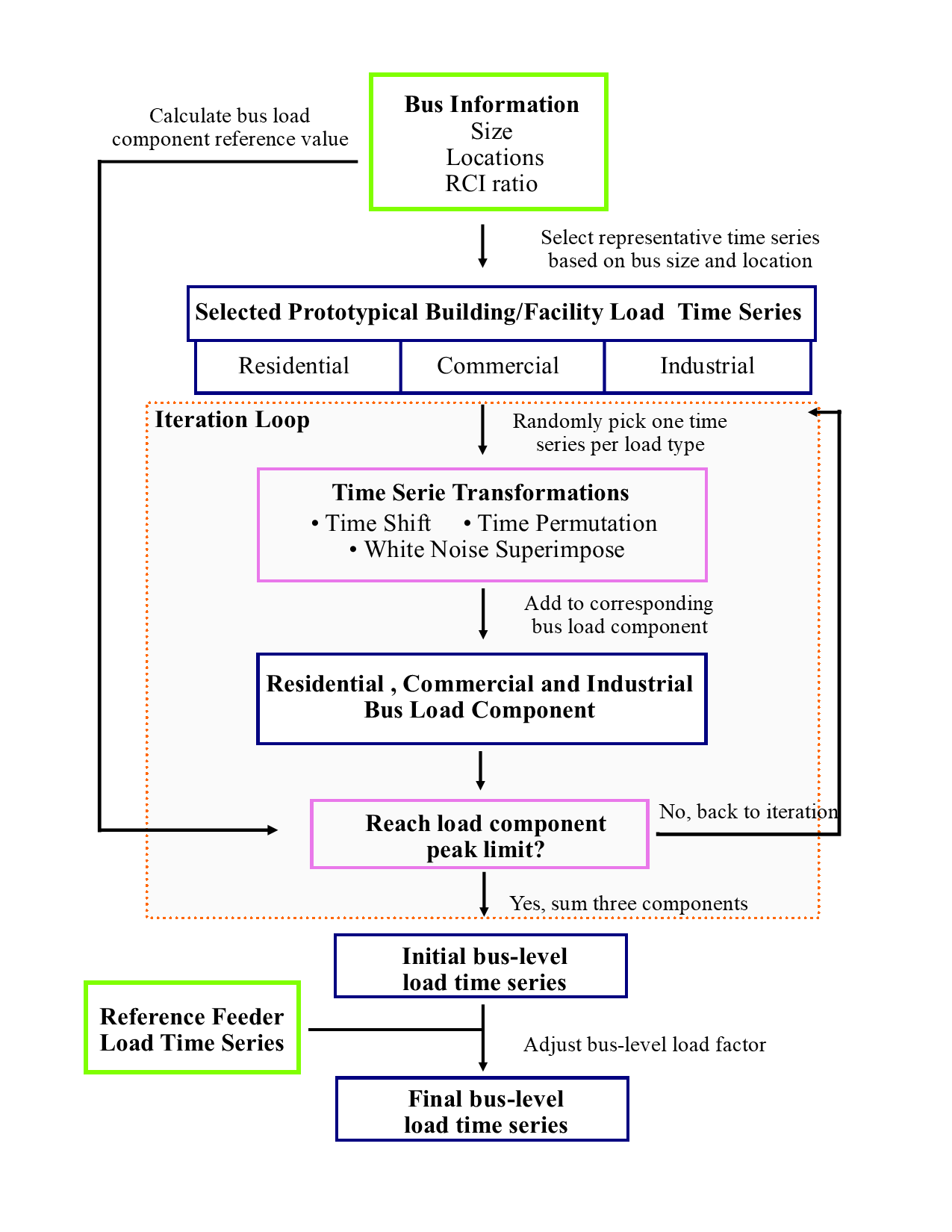}
  \caption{Flow chart of heuristic load aggregation approach}
  \label{fig:heuristic}
\end{figure}

A flow chart of this aggregation process is shown in figure \ref{fig:heuristic}. The reference peak values of residential, commercial and industrial bus load type of each bus are first determined by the multiplication of bus load size and the load composition ratio. This is used to integrate realistic amount of end users under each load type, where the peak component values are the stopping criterion for the iterative aggregation process. 

A pool of representative prototypical building load time series are then selected for each load bus. For residential and commercial load, the selection used the top five shortest distances between the load bus geographic coordinates and TMY3 locations. All industrial facility load time series that have smaller maximum value than the calculated peak industrial load component are included in this pool since industrial loads are less correlated with geographic locations. 

Under each load type, within one iteration, only one building or facility load time series is picked based on a heuristic weighted distribution as  equation (\ref{eq:1}), where time series with a lower average load is more likely to be chosen.

\begin{equation}
\label{eq:1}
P(TS_i) = \frac{\frac{1}{\sqrt{\overline{TS_i}}}}{\sum_{i=1}^{N} \frac{1}{\sqrt{\overline{TS_i}}}}
\end{equation}
where:

\begin{eqexpl}[13mm]
  \item{$P(TS_i)$}    probability of load time series i being chosen  \\
  \item{$N$}   number of time series selected in one load type\\ 
 \item{$\overline{TS_i}$}  yearly average value of load time series i \\   
\end{eqexpl}

 This selected building- or facility-level load time series is then processed through three types of transformations: time shift, time permutation, and small white noise insertion. Those transformations diversify the load profiles of end users, so that the smoothing effect for load aggregation can be produced. 

The original prototypical load time series can be shifted for an integer hour following a normal distribution in equation (\ref{eq:2}) , where the expectation is 0, and the standard deviation is determined based off the load type. 

\begin{equation}
\label{eq:2}
P(time \: shift = x) = \int_{x}^{x+1} \frac{1}{{\sigma \sqrt {2\pi } }}e^{{{ - \left( {t} \right)^2 } \mathord{\left/ {\vphantom {{ - \left( {t} \right)^2 } {2\sigma ^2 }}} \right. \kern-\nulldelimiterspace} {2\sigma ^2 }}} dt
\end{equation}
where:
\begin{eqexpl}[13mm]
 \item{$\sigma = 0.4$}    prototypical load time series is residential  \\
 \item{$\sigma = 0.3$}    prototypical load time series is commercial  \\
 \item{$\sigma = 0.1$}    prototypical load time series is industrial  \\
\end{eqexpl}

To imitate the random surges or drops of load for individual customers, certain hours (100, 100 and 50 hour pairs for residential, commercial and industrial respectively) are randomly chosen within the year to be permutated. To avoid bus load time series being overly conforming due to the use of similar prototypical building or facility time series, a small white Gaussian noise is also imposed. This transformed time series is then added to the corresponding type of load component, and the iteration would stop once after the load component maximum value calculated in previous step has been reached.

To mimic the effect of increasing load factor as load aggregates to a higher level \cite{AggregationEffect}, public feeder load time series managed by National Renewable Energy Laboratory (NREL) \cite{TaxonomyFeeders} is used to adjust the bus-level load factors to a realistic range. This distribution feeder load time series is populated from the taxonomy distribution feeders of different geographic regions.

The geographic coordinates of each load bus in the synthetic system are used to randomly select a subset of taxonomy feeders from the same geographic region, so that the summation of feeder load time series is on the same scale as the bus load. The load factor of the aggregated feeder load time series is calculated to be the reference value. A constant  component is added to the created bus-level load time series to adjust its load factor to a realistic value according to equation (\ref{eq:3}).

\begin{equation}
\label{eq:3}
\frac{Constant + Average\:Load}{Constant + Max\:Load} = Reference\:Load\:Factor
\end{equation}

\section{Example Results}
The load time series created for ACTIVSg2000 and ACTIVSg10K synthetic power systems are discussed in this paper as example results.  There are 1125 and 4170 load buses in those two cases, with 71 and 132 GW of system peak load respectively. 

\begin{figure}[!h]
  \centering
  \includegraphics[scale = 0.37]{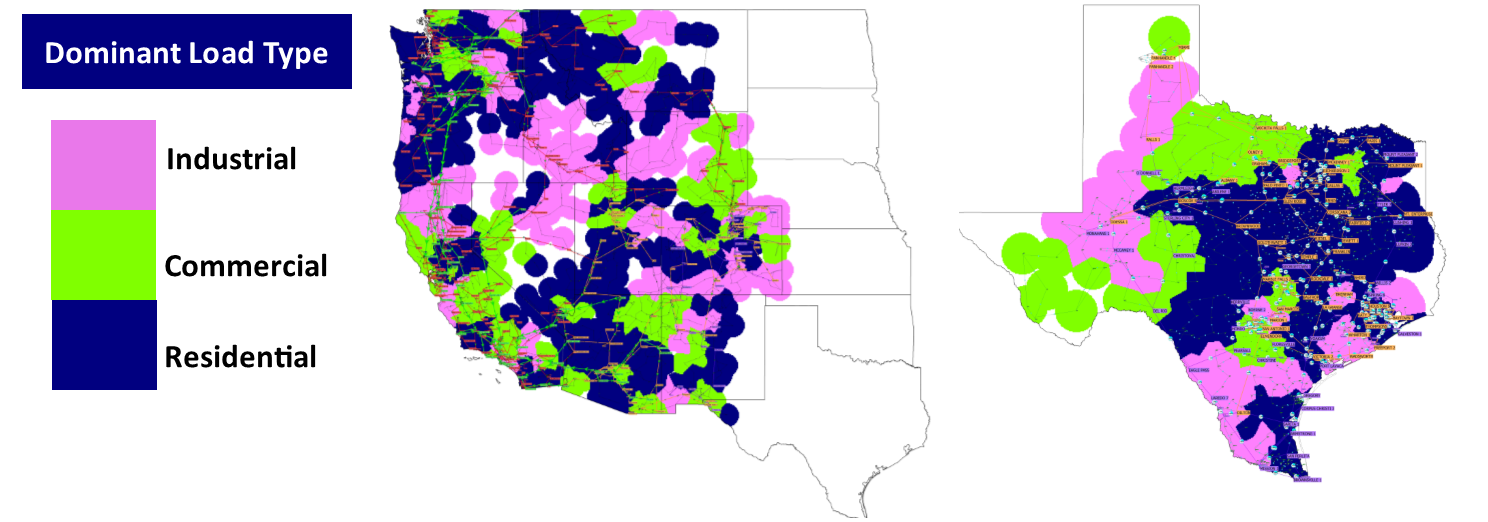}
  \caption{Dominant load type contour for ACTIVSg synthetic systems }
  \label{fig:2k_contour}
\end{figure}

The plots in figure \ref{fig:2k_contour} show the dominant bus load type using the method discussed in section II.B, where the load type with highest composition percentage is considered to be the dominant type of its load bus.  In ACTIVSg2000 system, 67.4\% of the buses are dominated by residential load, 18.8\% are primary composed of commercial load, and 13.8\% for industrial load. For the ACTIVSg10K system, 45.4\%, 33.7\% and 20.9\% of buses have residential, commercial and industrial load as primary composition respectively.

\begin{figure}[!h]
  \centering
  \includegraphics[scale = 0.4]{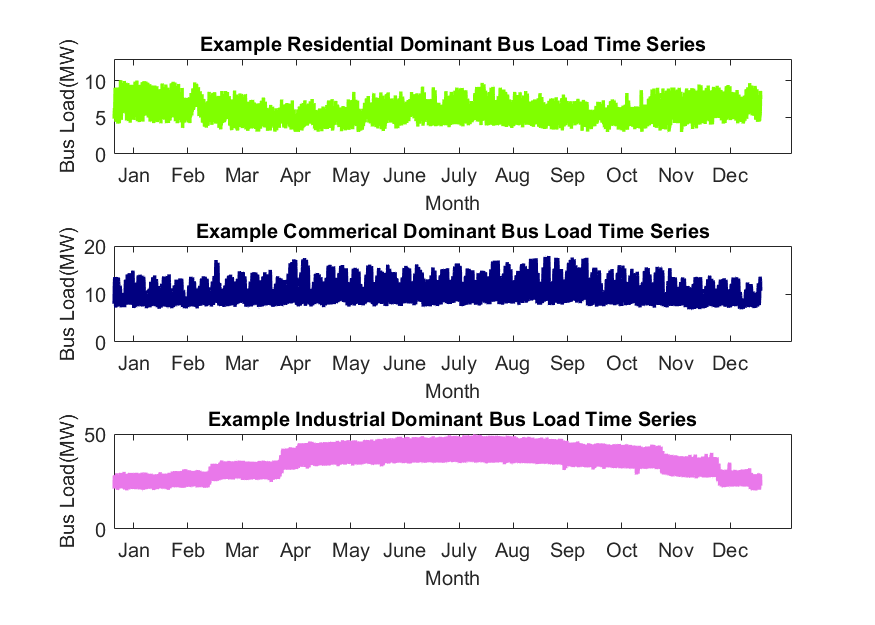}
  \caption{Bus-level synthetic load time series of different dominant load type}
  \label{fig:bus_load_result}
\end{figure}

On the bus-level, each load time series is unique based off the location and load composition ratio of the load bus. Average bus-level load time series of different dominant load types are shown in figure \ref{fig:bus_load_result}. Residentially-dominated bus load time series exhibit noticeable seasonal differences, where the electricity consumption in summer and winter seasons tend to have higher average values as well as higher variations. Commercially-dominated bus load time series have distinct daily patterns, while the electricity consumption base line stays relatively constant throughout the year.  Industrially-dominated bus loads usually have the lowest variation and highest load factor. The average size of industrially-dominated buses are larger than the other two types.

\begin{figure}[!h]
  \centering
  \includegraphics[scale = 0.4]{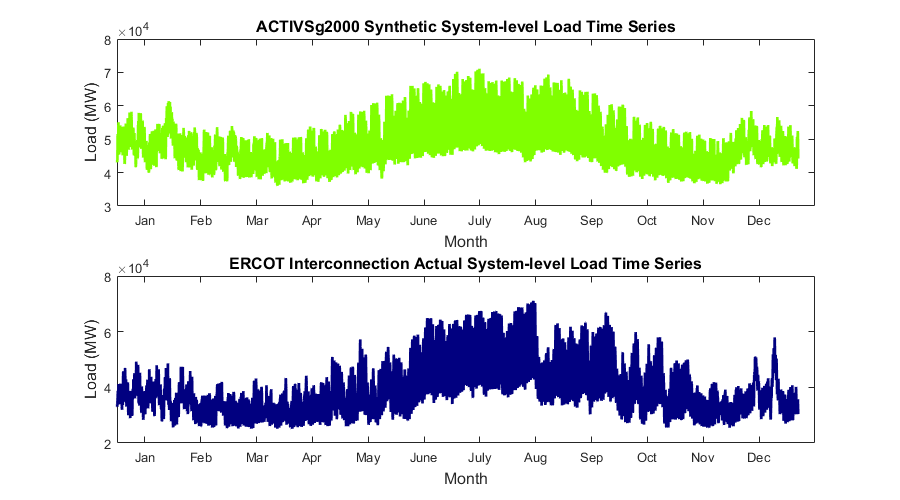}
  \caption{ACTIVSg2000 synthetic system load V.S. ERCOT Interconnection system Load}
  \label{fig:2k_system}
\end{figure}

\begin{figure}[!h]
  \centering
  \includegraphics[scale = 0.4]{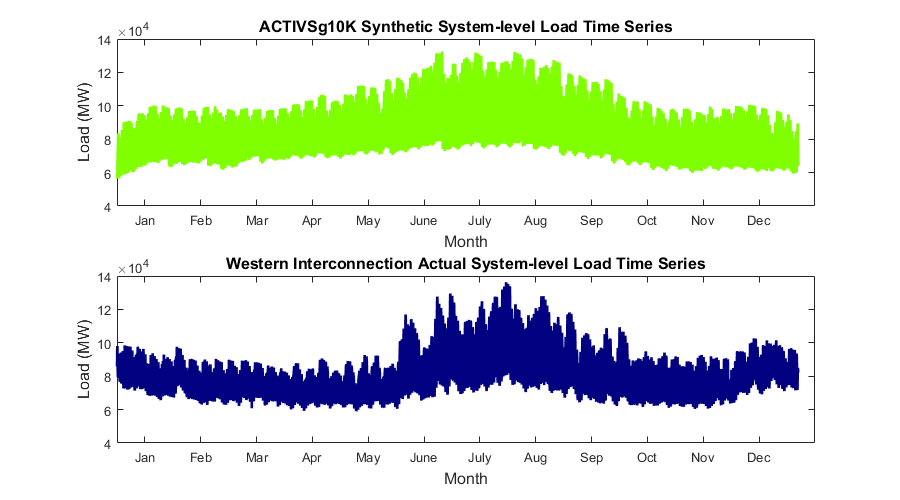}
  \caption{ACTIVSg10K synthetic system load V.S. Western Interconnection system load}
  \label{fig:10k_system}
\end{figure}

The system-level synthetic load time series and the actual system load from their footprint regions are shown in figure \ref{fig:2k_system} and figure \ref{fig:10k_system}. Although duplicating system-level load time series is not the desired outcome, synthetic load time series on the system level should exhibit similar general load shapes and trends compared to the actual system. 

Figure \ref{fig:2k_system} and figure \ref{fig:10k_system} show that the synthetic loads share similar size with the load of the actual system in the corresponding service territory. ACTIVSg2000 synthetic system has 71.1 GW of peak load, and 48.7 GW of average load, and the actual load of ERCOT system has 71.2 GW of maximum load and 41.0 GW of average load. For ACTIVSg10K system, there is 132.5 GW of peak load and 88.1 GW of average load. The corresponding actual system, the United States Western Interconnection, has 136.2 GW of peak load and 83.8 GW of average load.

Daily and weekly patterns can be seen from the ACTIVSg2000 and ACTIVSg10K synthetic load time series. It is also observed that the synthetic system has similar seasonal trends compared to the actual system. ACTIVSg2000 and ERCOT systems both experience peak load in summer, and also have some high load days weeks in winter. ACTIVSg10K and WI system also peak in summer, while their profile in the winter season is flatter.

\section{Validation of Synthetic Time Series}
since synthetic time series are fictitious, the validation of the created data against the actual data is critical to determine the quality and realism of the time series. A statistical based validation approach is implemented in this paper, as synthetic time series aim to realistically represent behaviors of load over time, instead of being an exact duplicate or forecast of the actual system time series. 

 A comprehensive set of validation metrics enables researchers to use synthetic time series with ease, but at the same time to be aware of the underlying assumptions. 

Validation metrics that are generic and independent from geographic locations are summarized using statistical characteristics found in public load data of  37 European countries \cite{OpenPowerSystemData} and 66 United States Balancing Authorities \cite{ElectricSystemOperationData}, so that synthetic load time series without a geographic footprint or have no availability of actual load time series can also be validated. 

\subsection{Load factors}
Load factor is defined as the ratio of average and peak value of a load time series. It is one effective metric to quantitatively validate the overall shape of the synthetic load profile. For profiles with relatively constant load level, such as regions with a high industrial composition, load factors are usually higher; while heavily residential areas tend to have lower load factors due to light occupation during the day \cite{LoadFactor}. 

\begin{figure}[!h]
  \centering
  \includegraphics[scale = 0.5]{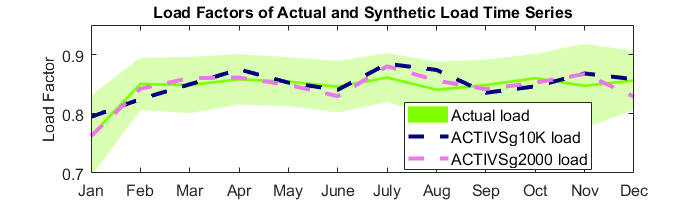}
  \caption{Monthly load factors of actual and synthetic load time series}
  \label{fig:BA_lf}
\end{figure}

The range of load factors of each month is summarized from public load data mentioned above and shown in figure \ref{fig:BA_lf} as the green shaded region. It is observed that as a general trend, the value of load factors are slightly higher in summer months, due to the increase of base electricity consumption from spacing cooling. There is also a consistent difference between the lowest and highest load factors of actual load every month, where systems with smaller size and less industrial load often have lower load factors. 

The load factors of ACTIVSg10K and ACTIVSg2000 load time series lie inside the range observed from actual load time series, and also follow the same monthly trend.

\subsection{Load distribution curves}
Load distribution curves show the percentage of time that load is at different levels. The load time series is normalized based off its mean value, where load levels exceeding yearly average would have per unit values larger than one.     

\begin{figure}[!h]
  \centering
  \includegraphics[scale = 0.5]{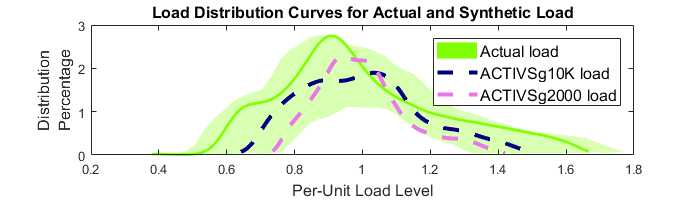}
  \caption{Load distribution curve validation}
  \label{fig:load_disrtibution}
\end{figure}

The green shaded band in Figure \ref{fig:load_disrtibution} shows the range of load distribution curves found in real load time series, where load levels are scattered in between 0.4 and 1.8 per unit, with a denser distribution in the range from 0.8 to 1.2. The distribution of ACTIVSg10K and ACTIVSg2000 load time series follow the same general trend, load at most of the time points are within 0.8 to 1.2 times its yearly average.

\subsection{Autocorrelations}
Autocorrelation exhibits the relationship between time points of load time series that are certain time lags apart. It provides a validation perspective in time sequence order, instead of observing time series values as if they are independent recordings.  

\begin{figure}[!h]
  \centering
  \includegraphics[scale = 0.5]{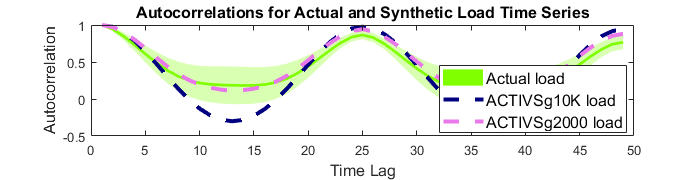}
  \caption{Load autocorrelation validation}
  \label{fig:load_autocorr}
\end{figure}

Figure \ref{fig:load_autocorr} shows the autocorrelations of actual and ACTIVSg synthetic load time series for time lags up to 48 steps. According to the real load time series data, the autocorrelation plot appears to be a sinusoidal-like wave of 24-hour cycle, with its magnitude slightly decreasing every cycle. The plot of ACTIVSg2000  synthetic load's autocorrelation exhibits a similar trend as the actual system, where  autocorrelation drops from 1 to around 0 when time lag increases from 0 to 12. Also, it increases back from 0 to 1 when time lag future increase from 12 to 24. The ACTIVSg10K synthetic load's autocorrelation also shares the same sinusoidal trend as real load time series's, while the lowest autocorrelation value would become negative around 12 hour time lag. This discrepancy is potentially introduced from the time shifting of prototypical building load time series during the load aggregation step. 

\section{Synthetic Time Series Applications Example}
The creation of synthetic time series enables the potential of scenario development, time sequence simulations, and wide-area visualizations of large-scale power systems. Those time series data reflect the typical behavior of electric load in a hourly manner for the whole year, which is fundamental for power system steady-state analysis desired at different times and of different durations. The buses each load time series locates also cover a wide range of North America regions and have longitudes and latitudes. This allows non-uniform alterations of the time series to construct realistic power system scenarios. The alterations can be made considering the coupling of power system and location- based factors such as weather and major events. Along with transmission system models, generator cost functions, and other system data, synthetic load time series can facilitate power flow, optimal power flow analysis, as well as unit commitment of various scenarios. 

As an example, this paper utilizes the bus-level load time series in the ACTIVSg10K system as the benchmark and develops a high behind-the-meter (BTM) solar scenario in the Western United States region. The BTM solar energies are solar generating units on the consumer's side of the retail meter that serve all or part of the customer's retail load with electric energy \cite{BTMG}. They are often treated as "negative loads" to be subtracted from the total load at the demand side.  The increasing capacity of BTM solar installation changes the shape of daily net load on the system-level, when solar generation peaks at midday, the net load is low and when solar generation trails off at the end of the day, the total demand ramps quickly upward \cite{DuckCurve}. This new load shape is often referred as a "duck curve". It has raised concerns on a conventional power system's ability to accommodate the ramp rate and range needed to effectively supply the load and fully utilize the renewable energy \cite{NRELDuckChart}.  

The bus-level load time series in the synthetic power system is utilized to construct rare and extreme scenarios that can be useful to study this impact over time. Based off the composition of bus load type and the location-based solar potentials, the benchmark load time series at each bus is altered so that a system-level "duck curve" is created for the ACTIVSg10K system. This "duck curve" scenario can be used as the input of power flow analysis and unit commitment to provide analytical insights on transmission line loadings, generator dispatches schedules, system costs, and other system conditions. 

This example uses a 24-hour time period in late spring from the benchmark hourly time series to develop a daily duck curve since such scenario usually occurs during the spring and summer seasons \cite{DuckCurve}. The BTM solar generation capacity is set to be 30,000 MW in the ACTIVSg10K system, and is distributed among load buses weighting their load sizes and the documented average solar resource outputs. Since most BTM solar installations are in the non-industrial sectors, only the size of residential and commercial load on each bus are considered to calculate the weights indicating bus solar potential, where buses with higher combined load are assigned with a higher peak BTM solar generation. On the other hand, as the solar potential is also dependent on solar radiance that varies with geographic locations, a solar resource data set from National Renewable Energy Laboratory (NREL) is also utilized to determine the weight to distribute the system BTM solar capacity \cite{NRELGIS}. This data set provides the monthly average solar output $(kWh/m^2/day)$ at each zip code location.  

The solar potential that determines the BTM solar capacity at each bus is calculated as the weighted summation of normalized load size and normalized solar resource output in equation (\ref{eq:4}) - (\ref{eq:5}). To construct the 24-hour "duck curve" scenario in the ACTIVSg10k synthetic system, a BTM solar output time series is created and subtracted from the original load time series of each bus. The bus-level BTM solar capacity is considered as the peak value of each BTM solar output time series that would occur at a random time step in between 1 pm and 3 pm.  The starting and ending time step of solar output are randomly chosen from 6 to 8 am, and 6 to 8 pm respectively. Before the starting and after the ending time point, the BTM solar output is zero.

\begin{equation}
\label{eq:4}
\begin{split}
BTM \: solar \: potential(i) & = X_1  \frac{load\:size(i)}{max(load\:size)} \\
& + X_2  \frac{solar\:resource(i)}{max(solar\:resource)}
\end{split}
\end{equation}

\begin{equation}
\label{eq:5}
\begin{split}
BTM \: solar \: capacity(i) & = system \: BTM \: solar \: capacity  \times \\
&  \frac{BTM \:solar \: potential(i)}{\sum_{i = 1}^{Nload}BTM \:solar \: potential(i)}
\end{split}
\end{equation}

where:

\begin{eqexpl}[20mm]
\item{$X_1,X_2$}    weights on bus load size and solar resource \\
\end{eqexpl}

\begin{figure}[!h]
  \centering
  \includegraphics[scale = 0.45]{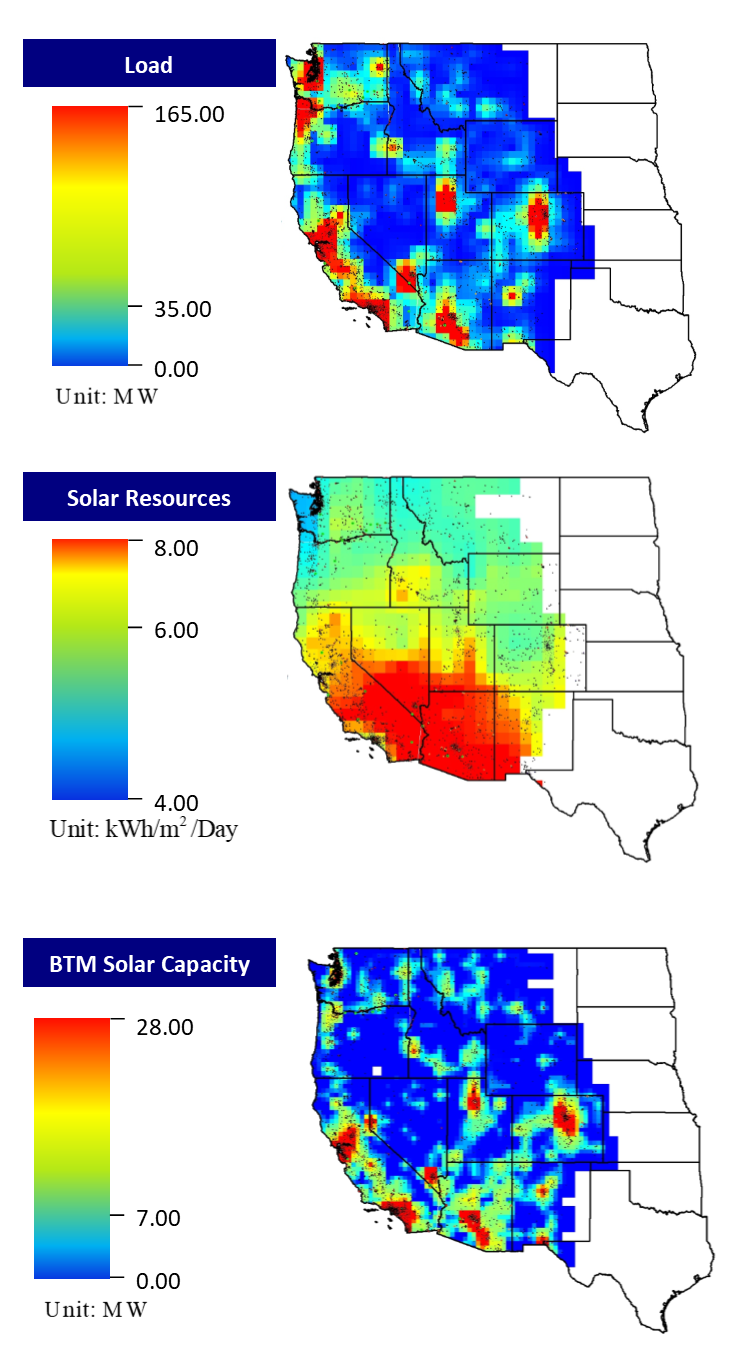}
  \caption{Contours of bus load, average solar resource, and BTM solar capacity in ACTIVSg10K synthetic system}
  \label{fig:contour}
\end{figure}

\begin{figure}[!h]
  \centering
  \includegraphics[scale = 0.45]{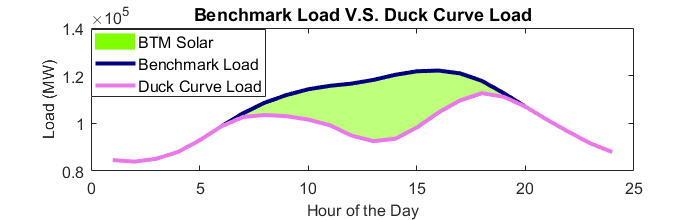}
  \caption{Benchmark load and duck curve load for ACTIVSg10K synthetic system}
  \label{fig:duckcurve}
\end{figure}

The load size and solar resources data, as well as the BTM solar capacity determined for buses in ACTIVSg10K system are shown in the contour plots Figure \ref{fig:contour}. It can be observed that most of the buses with higher BTM solar capacity are locations with high solar resources. Besides, major metropolitan areas with dense residential and commercial demands also have higher BTM solar capacities. The benchmark and duck curve load time series on the system-level are shown in Figure \ref{fig:duckcurve}.
 
\section{Conclusions}
This paper proposed a methodology to synthesize and validate bus-level load time series in the existing synthetic power systems. The creation of time series uses an iterative bottom-up approach. Based on the geographic location and load type composition of each bus, prototypical building and facility level time series are integrated to construct a bus-level time series with unique profiles. Each time series has hourly resolution, and spans for a year. To validate and improve the realism and quality of synthetic load time series, actual load time series obtained from electric systems of different sizes are analyzed statically so that representative and comprehensive set of validation metrics can be developed.

Since the data set utilized in the synthesizing process is publicly available, the created time series can be accessed and distributed freely without any confidentiality concerns. The wide geographic coverage, time resolution and duration of bus-level time series enable its versatile applications in system scenario development and studies. As an example, this paper demonstrated the construction of a "duck curve" scenario in ACTIVSg10K system using the bus-level load time series as the benchmark.

\section*{Acknowledgment}

This work was supported in part by the U.D. Department of Energy Advanced Research Projects Agency-Energy (ARPA-E) under the GRID DATA project.

\ifCLASSOPTIONcaptionsoff
  \newpage
\fi

\bibliographystyle{IEEEtran}
\bibliography{citation}

\renewenvironment{IEEEbiography}[1]
  {\IEEEbiographynophoto{#1}}
  {\endIEEEbiographynophoto}

\begin{IEEEbiography}{Hanyue Li}(S`14) received the B.Sc. degree in electrical
engineering from Illinois Institute of Technology, Chicago, IL,
USA, in 2016, and the M.Sc. degree in electrical and computer
engineering in Carnegie Mellon University, Pittsburgh, PA,
USA, in 2017. She is currently working toward the Ph.D.
degree in electrical engineering at Texas A\&M University,
College Station, TX, USA. 
\end{IEEEbiography}

\begin{IEEEbiography}{Ju Hee Yeo} (S`17) received the B.Sc. degree in electrical engineering from Sangmyung University, Seoul, South Korea, in 2017. She is currently working toward the Ph.D. degree in electrical engineering at Texas A\&M University,
College Station, TX, USA. 
\end{IEEEbiography}

\begin{IEEEbiography}{Ashly L. Bornsheuer} (S`17) received the B.Sc. degree in electrical engineering from
Texas A\&M University, College Station, TX, USA in 2018.
\end{IEEEbiography}

\begin{IEEEbiography}{Thomas J. Overbye} (S`87-M`92-SM`96-F`05) received the
B.Sc., M.Sc., and Ph.D. degrees in electrical engineering from
the University of Wisconsin-Madison, Madison, WI, USA. He
is currently a TEES distinguished research Professor in the
Department of Electrical and Computer Engineering at Texas
A\&M University, College Station, TX, USA.

\end{IEEEbiography}

\end{document}